\documentclass[12pt,a4paper]{article}

\usepackage{textcomp}
\usepackage[latin1]{inputenc}
\usepackage{pstricks,pst-node,pst-text}
\usepackage{amsmath}
\usepackage{graphicx}

\newcommand{\bc}{\begin{center}}
\newcommand{\ec}{\end{center}}
\newcommand{\be}{\begin{equation}}
\newcommand{\ee}{\end{equation}}
\newcommand{\bea}{\begin{eqnarray}}
\newcommand{\eea}{\end{eqnarray}}
\newcommand{\bal}{\begin{aligned}}
\newcommand{\eal}{\end{aligned}}
\newcommand{\ba}{\begin{array}}
\newcommand{\ea}{\end{array}}
\newcommand{\ben}{\begin{enumerate}}
\newcommand{\een}{\end{enumerate}}
\newcommand{\bitem}{\begin{itemize}}
\newcommand{\eitem}{\end{itemize}}

\newcommand{\fr}{\frac}

\newcommand{\crn}{\nonumber \\}

\newcommand{\noi}{\noindent}

\newcommand{\Ga}{\Gamma}

\newcommand{\eps}{\epsilon}

\newcommand{\epsp}{\eps^\prime}

\newcommand{\hs}{\hspace*{3mm}}
\newcommand{\ie}{{\it i.e.}}

\newcommand{\eq}[1]{Eq.~(\ref{#1})}
\newcommand{\Eq}[1]{Eq.~(\ref{#1})}
\newcommand{\fig}[1]{Fig.~\ref{#1}}
\newcommand{\tab}[1]{Table~\ref{#1}}
\newcommand{\sect}[1]{Section~\ref{#1}}
\newcommand{\ssect}[1]{Subsection~\ref{#1}}

\newcommand{\appen}[1]{Appendix~\ref{#1}}

\newcommand{\loga}{logarithm}
\newcommand{\logas}{logarithms}
\newcommand{\etaf}{$\eta$-function}
\newcommand{\betatr}{beta triangle}

\newcommand{\lt}{{\texttt{LoopTools}}}
\newcommand{\doc}{{\texttt{D0C}}}
\newcommand{\dor}{{\texttt{D0}}}
\newcommand{\ff}{{\texttt{FF}}}
\newcommand{\bases}{{\texttt{BASES}}}
\newcommand{\verslt}{{\texttt{LoopTools-2.4}}}

\newcommand{\iepsp}{$i\eps$-prescription}

\newcommand{\df}{scalar one-loop four-point function}
\newcommand{\dfi}{scalar one-loop four-point integral}
\newcommand{\tti}{\texttt{i}}
\newcommand{\dcn}{\texttt{D0Cn.F}}
 
\newcommand{\dcabc}{\texttt{d0c\_abc.F}}
\newcommand{\ftn}{\texttt{fTn.F}} 
\newcommand{\ftnresd}{\texttt{fTn\_resd.F}} 
\newcommand{\ftnlin}{\texttt{fTn\_lin.F}} 
\newcommand{\ftot}{\texttt{fT13.F}}
\newcommand{\fsn}{\texttt{fS3n.F}}
\newcommand{\fstwo}{\texttt{fS2.F}}
\newcommand{\frf}{\texttt{fR.F}}

\newcommand{\tf}{\mathcal{T}}
\newcommand{\rf}{\mathcal{R}}
\newcommand{\nf}{\mathcal{N}}

\newcommand{\sfun}{\mathcal{S}}
\newcommand{\Rel}{{\rm{Re}}}
\newcommand{\Img}{{\rm{Im}}\,}
\newcommand{\sign}{{\rm{sign}}}

\def\slashepi{\epsilon_i\kern -.720em {/}}
\def\slashpi{p_i\kern -.600em {/}}

\begin{document}

\begin{titlepage}

\vspace*{0.1cm}\rightline{MPP-2009-16}


\vspace{1mm}
\begin{center}

{\Large{\bf \doc : A code to calculate scalar one-loop four-point integrals with complex masses}}

\vspace{.5cm}

DAO Thi Nhung and LE Duc Ninh

\vspace{4mm}

{\it Max-planck-Institut f\"ur Physik (Werner-Heisenberg-Institut), \\
D-80805 M\"unchen, Germany}

\vspace{10mm}
\abstract{We present a new Fortran code 
to calculate the scalar one-loop 
four-point integral with complex internal masses, 
based on the method of 
't Hooft and Veltman. The code is applicable when 
the external momenta fulfill a certain physical condition. 
In particular it holds if one of the external momenta or a 
sum of them is timelike or lightlike and 
therefore covers all physical processes at colliders. 
All the special cases 
related to massless external particles are treated separately. 
Some technical issues related to numerical evaluation and Landau singularities 
are discussed.}
\end{center}

\normalsize

\end{titlepage}

\section{Introduction}
In order to calculate the radiative corrections to reactions with 
unstable internal particles (like the top quark or W/Z gauge bosons) 
which can be on-shell, one has to resum the propagator of these particles. 
This can lead to prescriptions where the squared mass of unstable particles are
regarded as complex parameters with a 
finite imaginary part proportional to the width. 
It means that we have to calculate Feynman integrals with complex 
masses. 

From the mathematical viewpoint, the introduction of a width helps 
to protect the cross section from several singularities, called Landau 
singularities \cite{landau, book_eden, ninh_bbH2, Goria:2008ny, Actis:2008uh, ninh_thesis}. Let us explain this in a simple way. In order to 
define a Feynman integral in the case of real masses, one needs the \iepsp\ 
to deform the integration contour from various poles. 
The final result is obtained by taking the limit $\eps\to 0$. 
We observe that the position of the poles, in general, depend on 
the values of internal masses and external momenta (see \appen{appen_real} for more details). It is therefore possible 
that these poles can pinch the integration contour in the limit $\eps\to 0$, leading 
to a Landau singularity. 
By making the masses complex, the poles are moved away from the real axis 
and hence all the various Landau singularities no longer exist.

The method to calculate scalar one-loop integrals with real/complex masses was 
explained by 't Hooft and Veltman
\cite{hooft_velt}. Following this, van Oldenborgh has written 
a Fortran package \ff\ to calculate one-loop integrals \cite{ff, ff0}. 
Based on this package, a more convenient library called \lt\ which 
can calculate scalar and tensor one-loop integrals up to five-point has 
been developed by Hahn and colleagues \cite{looptools, looptools_5p}. 
\lt /\ff\  can handle one-loop integrals with complex 
masses up to three-point. The only missing piece in the scalar integral 
sector is the one-loop four-point function with complex 
masses. A Fortran 77 code to calculate this function is presented in this 
paper. The method is based on the one given in \cite{hooft_velt} where 
a general result is written down, see Eq.~(6.26) of \cite{hooft_velt}. 
This result is applicable as long as the external momenta fulfill a certain physical condition. 
In particular it holds if one of the external momenta or a 
sum of them is timelike\footnote{In our metric timelike $p$ implies positive $p^2$.}.
If there is at least one lightlike external 
momentum then similar results can be derived by using the same method.  
If the condition is not met then the calculation becomes mathematically quite tricky. 
In principle, the result in this case should be 
an analytical continuation of Eq.~(6.26) of \cite{hooft_velt} with 
some extra logarithms. However, it is not easy to find these terms in practice. 
Fortunately, this special case is never encountered in collider scatterings where 
there always exists at least two timelike/lightlike external momenta. 

The outline of this paper is as follows. In the next section, we give all 
the relevant formulae which have been implemented in our code, named \doc . All the important facts related to the code are given in \sect{code}. 
The appendices include a discussion on three-point and four-point Landau singularities and the 
calculation of a basic integral.

\section{The method}
The notation used in this paper follows closely the one of Ref.~\cite{hooft_velt}. 
The definition of \loga , Spence function and \etaf\ 
are the same. The signature of our space-time metric is $(+,-,-,-)$ which is different from 
the one used by 't Hooft and Veltman, however.
\begin{figure}[hbt]
\begin{center}
\includegraphics[width=6cm]{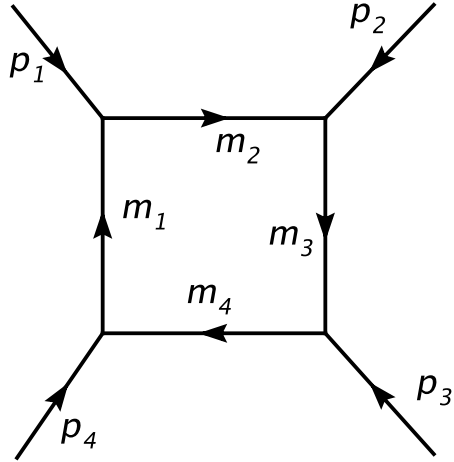}
\caption{\label{box_diag}{\em The box diagram.}}
\end{center}
\end{figure}

With the momenta assigned as in \fig{box_diag}, the scalar 
one-loop four-point function in 4-dimension space time is defined by
\bea
D_{0}&=&\int\fr{d^4q}{i\pi^2}\fr{1}{\prod_{j=1}^4[(q+k_j)^2-m_j^2+i\eps]}\label{d0_def_ki}
\eea
where the momenta $k_i$ are related to the {\em real} external momenta as
\bea
k_1=0,\hs k_2=p_1,\hs k_3=p_1+p_2, \hs k_4=p_1+p_2+p_3
\eea
and the internal masses can have a {\em non-positive} imaginary part
\bea
m_i^2=m_i^2-im_i\Ga_i, \hs m_i\Ga_i\ge 0, \hs i=1,2,3,4
\eea
with $\Ga_i$ are the width of the internal particles.  
The Feynman parameter
representation reads
\bea
D_{0}&=&\int_0^\infty dx_1dx_2dx_3dx_4\fr{\delta(\sum_{i=1}^{4}x_i-1)}{\Delta^2},\crn
\Delta&=&\fr{1}{2}\sum_{i,j=1}^4x_ix_jQ_{ij}-i\eps,\hs Q_{ij}=m_i^2+m_j^2-(k_i-k_j)^2
\label{d0_def_xi}.
\eea
$Q_{ij}$ is called the Landau matrix. By changing the 
integration variables as $t=\sum_{i=1}^4x_i$, $x=\sum_{i=1}^3x_i$, $y=x_1+x_2$, $z=x_1$ 
and integrating out the Dirac delta function we obtain
\bea
D_0&=&\int_0^1dx\int_0^xdy\int_0^ydz\crn
&\times&\fr{1}{(ax^2+by^2+gz^2+cxy+hxz+jyz+dx+ey+kz+f-i\eps)^2}\label{d0_xyz}
\eea
with
\bea
\bal
a&=p_3^2,\hs b=p_2^2,\hs g=p_1^2,\hs c=2p_2.p_3,\hs h=2p_1.p_3,\hs j=2p_1.p_2,\\
d&=m_3^2-m_4^2-p_3^2,\hs e=m_2^2-m_3^2-p_2^2-2p_2.p_3,\hs k=m_1^2-m_2^2+p_1^2+2p_1.p_4,\\
f&=m_4^2.
\eal
\eea
It is important to notice that $a$, $b$, $g$, $c$, $h$ and $j$ are real while $d$, $e$, $k$ and 
$f$ are complex. The denominator always has a negative imaginary part as indicated by the "$-i\eps$"
\bea
\Img (dx+ey+kz+f-i\eps)=\Img [zm_1^2+(y-z)m_2^2+(x-y)m_3^2+(1-x)m_4^2-i\eps]<0.
\eea
The Landau determinant as a function of new variables reads
\bea
\bal
\det(Q)= &\  k^2 c^2+e^2 h^2+d^2 j^2-2 e h kc-2 d jkc-2 d e h j\\
&+4cd e g+4cf h j-4 f gc^2-4 a f j^2-4 a e^2 g+4 ae j k\\
&-4 b g d^2+4 bh k d-4 bf h^2-4 ba k^2+16ba f g.
\eal
\label{detQ_abc}
\eea
The \dfi\ is a function of 10 variables:
\bea
D_0=D_0(a_1,a_2,a_3,a_4,a_5,a_6;b_1,b_2,b_3,b_4)
\eea
where the set of variables are given in \tab{sym_d0} which shows 
all the symmetries of this function\footnote{\Eq{d0_def_xi} shows that 
$D_0$ has $4!=24$ symmetries.}.
\begin{table}[h]
\small
\bc
\caption{Symmetries of $D_0$}
\begin{tabular}{cccccccccc}
	\hline\hline
$a_1$ &  $a_2$ & $a_3$ & $a_4$ & $a_5$ & $a_6$ & $b_1$ & $b_2$ & $b_3$ & $b_4$  \\
        \hline\hline	
$p_1^2$ &  $p_2^2$ & $p_3^2$ & $p_4^2$ & $(p_1+p_2)^2$ & $(p_2+p_3)^2$ & $m_1^2$ & $m_2^2$ & $m_3^2$ & $m_4^2$ \\
\hline	
$p_1^2$ &  $(p_2+p_3)^2$ & $p_3^2$ & $(p_1+p_2)^2$ & $p_4^2$ & $p_2^2$ & $m_1^2$ & $m_2^2$ & $m_4^2$ & $m_3^2$ \\
\hline	
$(p_1+p_2)^2$ & $p_2^2$ & $(p_2+p_3)^2$ & $p_4^2$ & $p_1^2$ & $p_3^2$ & $m_1^2$ & $m_3^2$ & $m_2^2$ & $m_4^2$ \\
\hline	
$(p_1+p_2)^2$ &  $p_3^2$ & $(p_2+p_3)^2$ & $p_1^2$ & $p_4^2$ & $p_2^2$ & $m_1^2$ & $m_3^2$ & $m_4^2$ & $m_2^2$ \\
\hline	
$p_4^2$ &  $(p_2+p_3)^2$ & $p_2^2$ & $(p_1+p_2)^2$ & $p_1^2$ & $p_3^2$ & $m_1^2$ & $m_4^2$ & $m_2^2$ & $m_3^2$ \\
\hline	
$p_4^2$ &  $p_3^2$ & $p_2^2$ & $p_1^2$ & $(p_1+p_2)^2$ & $(p_2+p_3)^2$ & $m_1^2$ & $m_4^2$ & $m_3^2$ & $m_2^2$ \\
\hline	
$p_1^2$ &  $(p_1+p_2)^2$ & $p_3^2$ & $(p_2+p_3)^2$ & $p_2^2$ & $p_4^2$ & $m_2^2$ & $m_1^2$ & $m_3^2$ & $m_4^2$ \\
\hline	
$p_1^2$ &  $p_4^2$ & $p_3^2$ & $p_2^2$ & $(p_2+p_3)^2$ & $(p_1+p_2)^2$ & $m_2^2$ & $m_1^2$ & $m_4^2$ & $m_3^2$ \\
\hline	
$p_2^2$ &  $(p_1+p_2)^2$ & $p_4^2$ & $(p_2+p_3)^2$ & $p_1^2$ & $p_3^2$ & $m_2^2$ & $m_3^2$ & $m_1^2$ & $m_4^2$ \\
\hline	
$p_2^2$ &  $p_3^2$ & $p_4^2$ & $p_1^2$ & $(p_2+p_3)^2$ & $(p_1+p_2)^2$ & $m_2^2$ & $m_3^2$ & $m_4^2$ & $m_1^2$ \\
\hline	
$(p_2+p_3)^2$ &  $p_4^2$ & $(p_1+p_2)^2$ & $p_2^2$ & $p_1^2$ & $p_3^2$ & $m_2^2$ & $m_4^2$ & $m_1^2$ & $m_3^2$ \\
\hline	
$(p_2+p_3)^2$ &  $p_3^2$ & $(p_1+p_2)^2$ & $p_1^2$ & $p_2^2$ & $p_4^2$ & $m_2^2$ & $m_4^2$ & $m_3^2$ & $m_1^2$ \\
\hline	
$(p_1+p_2)^2$ &  $p_1^2$ & $(p_2+p_3)^2$ & $p_3^2$ & $p_2^2$ & $p_4^2$ & $m_3^2$ & $m_1^2$ & $m_2^2$ & $m_4^2$ \\
\hline	
$(p_1+p_2)^2$ &  $p_4^2$ & $(p_2+p_3)^2$ & $p_2^2$ & $p_3^2$ & $p_1^2$ & $m_3^2$ & $m_1^2$ & $m_4^2$ & $m_2^2$ \\
\hline	
$p_2^2$ & $p_1^2$ & $p_4^2$ & $p_3^2$ & $(p_1+p_2)^2$ & $(p_2+p_3)^2$ & $m_3^2$ & $m_2^2$ & $m_1^2$ & $m_4^2$ \\
\hline	
$p_2^2$ & $(p_2+p_3)^2$ & $p_4^2$ & $(p_1+p_2)^2$ & $p_3^2$ & $p_1^2$ & $m_3^2$ & $m_2^2$ & $m_4^2$ & $m_1^2$ \\
\hline	
$p_3^2$ &  $p_4^2$ & $p_1^2$ & $p_2^2$ & $(p_1+p_2)^2$ & $(p_2+p_3)^2$ & $m_3^2$ & $m_4^2$ & $m_1^2$ & $m_2^2$ \\
\hline	
$p_3^2$ &  $(p_2+p_3)^2$ & $p_1^2$ & $(p_1+p_2)^2$ & $p_2^2$ & $p_4^2$ & $m_3^2$ & $m_4^2$ & $m_2^2$ & $m_1^2$ \\
\hline	
$p_4^2$ &  $p_1^2$ & $p_2^2$ & $p_3^2$ & $(p_2+p_3)^2$ & $(p_1+p_2)^2$ & $m_4^2$ & $m_1^2$ & $m_2^2$ & $m_3^2$ \\
\hline	
$p_4^2$ &  $(p_1+p_2)^2$ & $p_2^2$ & $(p_2+p_3)^2$ & $p_3^2$ & $p_1^2$ & $m_4^2$ & $m_1^2$ & $m_3^2$ & $m_2^2$ \\
\hline	
$(p_2+p_3)^2$ &  $p_1^2$ & $(p_1+p_2)^2$ & $p_3^2$ & $p_4^2$ & $p_2^2$ & $m_4^2$ & $m_2^2$ & $m_1^2$ & $m_3^2$ \\
\hline	
$(p_2+p_3)^2$ &  $p_2^2$ & $(p_1+p_2)^2$ & $p_4^2$ & $p_3^2$ & $p_1^2$ & $m_4^2$ & $m_2^2$ & $m_3^2$ & $m_1^2$ \\
\hline	
$p_3^2$ &  $(p_1+p_2)^2$ & $p_1^2$ & $(p_2+p_3)^2$ & $p_4^2$ & $p_2^2$ & $m_4^2$ & $m_3^2$ & $m_1^2$ & $m_2^2$ \\
\hline	
$p_3^2$ &  $p_2^2$ & $p_1^2$ & $p_4^2$ & $(p_2+p_3)^2$ & $(p_1+p_2)^2$ & $m_4^2$ & $m_3^2$ & $m_2^2$ & $m_1^2$ \\
	\hline\hline
\end{tabular}
\label{sym_d0}
\ec
\end{table}
Using those symmetries, we see that the notation of external momentum 
is extended to include also $(p_1+p_2)$ and $(p_2+p_3)$. This definition 
of external momentum will be used hereafter.

\subsection{At least one lightlike momentum}
If there exists at least 
one lightlike momentum which we can always choose to be $p_1$ then the denominator 
in \eq{d0_xyz} becomes linear in $z$. The last integral can be easily performed to get
\bea
D_0^{(1)}=\tf(a,b,c,h,j;d,e,f,k)-\tf(a,b+j,c+h,h,j;d,e+k,f,k), \label{d0_1zero}
\eea
where the calculation of the $\tf$ function is given in \ssect{T-func}. The result 
contains $72$ Spence functions since each $\tf$ function contains $36$ Spence functions.

In the case where $D_0$ has two non-adjacent lightlike momenta, say $p_1^2=p_3^2=0$, 
the result can be written in terms of two $\tf$ functions as above. However, one should 
rather combine them into one single function as (in this way, the result contains 
a lesser number of Spence functions)
\bea
D_0^{(13)}&=&\int_0^1dx\int_0^xdy\crn
&\times&\fr{y}{(Gy^2+Hxy+Dx+Iy+F-i\eps)(Ay^2+Cxy+Dx+Ey+F-i\eps)}\label{d0_2zero}
\eea 
where
\bea
A=b+j,\hs C=c+h, \hs D=d, \hs E=e+k, \crn
F=f,\hs G=b, \hs H=c, \hs I=e.
\eea
We notice that each denominator is a linear function in $x$ hence one can perform the 
$x$-integral by using:
\bea
\int_0^1dx\int_0^xdy=\int_0^1dy\int_y^1dx.\label{int_x_y}
\eea
The result is written in terms of $32$ Spence functions \cite{ninh_bbH2, ninh_thesis}
\bea
D_0^{(13)}&=&\int_0^1\fr{dy}{SV-TU}
\left(\ln\fr{S+T}{U+V}-\ln\fr{Sy+T}{Uy+V}\right)\crn
&=&\fr{1}{(HA-CG)(y_2-y_1)}\sum_{i=1}^2\sum_{j=1}^4(-1)^{i+j}
\int_0^1\fr{dy}{y-y_i}\ln(A_jy^2+B_jy+C_j),\label{d0_y_13_sumij}
\eea
where 
\bea
S&=&Hy+D,\hs T=Gy^2+Iy+F-i\eps,\crn
U&=&Cy+D,\hs V=Ay^2+Ey+F-i\eps;
\eea
and $y_{1,2}$ are the two roots of the equation
\bea
(HA-CG)y^2+(AD+HE-DG-CI)y+DE-DI+(H-C)(F-i\eps)=0
\eea
or\footnote{In the case of real masses, the "$i\eps$"-part of each root 
should be written separately as shown in \eq{root_y12_eps}.}
\bea
y_{1,2}=\fr{-(AD+HE-DG-CI)\mp\sqrt{\det(Q)}}{2(HA-CG)}
\eea
where the indices $1$, $2$ correspond to $-$ and $+$ signs respectively and the 
discriminant,
\bea
\det(Q)=(AD+HE-DG-CI)^2-4(HA-CG)(DE-DI+HF-CF),
\eea
is nothing but the determinant of the Landau matrix $Q$ defined in \eq{detQ_abc}. The various 
coefficients of the argument of the logarithms are given in \tab{coeff_D0_13}. The integral in \eq{d0_y_13_sumij} 
will be considered in \appen{appen_T1}.
\begin{table}[h]
\bc
\caption{The coefficients of the $D_0^{(13)}$-function}
\begin{tabular}{cccc}
	\hline
$j$ &  $A_j$ & $B_j$ & $C_j$ \\
	\hline
$1$   & $A$ & $E+C$ & $D+F-i\eps$  \\
$2$   & $G$ & $I+H$ & $D+F-i\eps$  \\
$3$   & $G+H$ & $D+I$ & $F-i\eps$  \\
$4$   & $A+C$ & $E+D$ & $F-i\eps$  \\
	\hline
\end{tabular}
\label{coeff_D0_13}
\ec
\end{table}

\subsection{General case}
As proven in \cite{hooft_velt} (see Eq.~(6.18)), $D_0$ can be written as
\bea
D_0&=&-\tf(a+b+c,g,j+h,c+2b+h\alpha+j\alpha,j+2\alpha g;d+e,k,f,e+k\alpha)\crn
&+&(1-\alpha)\tf(a,b+g+j,c+h,c+h\alpha,(j+2\alpha g)(1-\alpha);d,e+k,f,e+k\alpha)\crn
&+&\alpha\tf(a,b,c,c+h\alpha,-(j+2\alpha g)\alpha;d,e,f,e+k\alpha)
\label{d0_T}
\eea
where $\alpha$ is a root of the equation
\bea
g\alpha^2+j\alpha+b=0.
\eea
The method is restricted to real $\alpha$, \ie
\bea 
j^2-4bg&=&4[(p_1\cdot p_2)^2-p_1^2p_2^2]\crn
&=&\lambda[(p_1+p_2)^2,p_1^2,p_2^2]\ge 0, \label{cond_alpha}\\
\lambda(x,y,z)&=&x^2+y^2+z^2-2xy-2xz-2yz.\eea 
This is satisfied if $p_1$(or $p_2$ or $p_1+p_2$) is timelike. If 
the above condition is not fulfilled then one has to keep rotating the external momenta by 
using the symmetric properties given in \tab{sym_d0} until it is met.
The right hand side (rhs) of \eq{d0_T} contains $36\times 3=108$ Spence functions.
\subsection{The $\tf$ function}
\label{T-func}
We now consider the following $\tf$ function
\bea
\tf(A,B,C,G,H;D,E,F,J)=\int_0^1dx\int_0^xdy\hspace*{5.6cm}\crn
\hspace*{2cm}\times\fr{1}{(Gx+Hy+J)(Ax^2+By^2+Cxy+Dx+Ey+F-i\eps)}\label{def_T0},
\eea 
where $A$, $B$, $C$, $G$ and $H$ are real and $D$, $E$, $F$ and $J$ can be complex with the 
restriction that the imaginary part of the second denominator has a definite sign, say negative as indicated 
by the $-i\eps$. In order to make the imaginary part of the first denominator also negative, 
we multiply both the numerator and denominator of the integrand with $-s_J=-\sign(J)$ and get
\bea
\tf=-s_J\int_0^1dx\int_0^xdy\fr{1}{(Gx+Hy+J-i\epsp)(Ax^2+By^2+Cxy+Dx+Ey+F-i\eps)}\label{def_T}
\eea
where $G=-s_JG$, $H=-s_JH$, $J=-s_JJ$ and $\epsp$ is infinitesimally positive. 

\subsubsection{General case: $A\neq 0$ and $B\neq 0$}
\begin{figure}[hbt]
\begin{center}
\includegraphics[width=6cm]{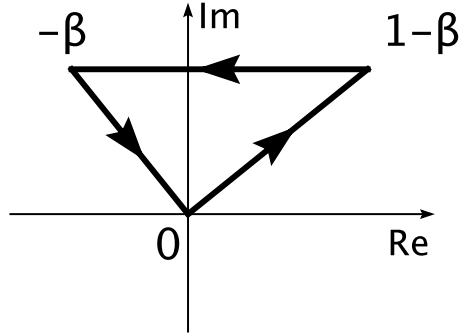}
\caption{\label{triangle_beta}{\em The \betatr\ $[0,1-\beta,-\beta]$ in 
the complex plane.}}
\end{center}
\end{figure}

Following the method explained in \cite{hooft_velt}, we obtain 
(similar to Eq.~(6.26) of \cite{hooft_velt})
\bea
-s_J\tf&=&\fr{1}{K(y_2-y_1)}\sum_{i=1}^2\sum_{j=1}^6(-1)^{i}\int_0^1dy\fr{c_j}{(-1)^{j+1}c_jy-b_j-y_i}
\ln(A_jy^2+B_jy+C_j)\crn
&-&\fr{1}{K(y_2-y_1)}\sum_{i=1}^2(-1)^{i}P_i\ln{R_i}
\label{T_result}
\eea
with the following cascade of notation.  
\bea
y_1=\fr{-L-\sqrt{\det(Q)}}{2K}, \hs y_2=\fr{-L+\sqrt{\det(Q)}}{2K}, \hs 
\det(Q)=L^2-4NK
\eea
are the roots and the discriminant of the equation
\bea
Ky^2+Ly+N&=&0, \hs K=B(G+\beta H)-H(C+2\beta B), \crn
L&=&(G+\beta H)E-H(D+\beta E)-J(C+2\beta B), \crn 
N&=&(G+\beta H)(F-i\eps)-(D+\beta E)J\eea
with $\beta$ is one root of the equation
\bea
B\beta^2+C\beta+A=0.
\label{eq_beta}
\eea
It is easy to check that the discriminant is free of $\beta$ 
and is indeed the determinant of the Landau matrix $Q$ if our calculation is set in the 
context of \eq{d0_1zero} or \eq{d0_T}. And
\bea
R_i&=&\fr{Hy_i+J}{By_i^2+Ey_i+F-i\eps}=\fr{G+\beta H}{(C+2\beta B)y_i+D+\beta E},\crn
P_i&=&\sum_{j=1}^3\int_0^1dy\fr{c_j}{(-1)^{j+1}c_jy-b_j-y_i}=\oint_{\triangle_\beta}dy\fr{1}{y-y_i},
\eea
where $\triangle_\beta$ is the \betatr\ $[0,1-\beta,-\beta]$ as shown in \fig{triangle_beta}.
The second term in \eq{T_result} is nothing but the residue contribution.
All the other coefficients are listed in \tab{coeff_T}.
\begin{table}[h]
\bc
\caption{The coefficients of the $\tf$ function}
\begin{tabular}{cccccc}
	\hline
$j$ &  $c_j$    &   $b_j$  & $A_j$ & $B_j$ & $C_j$ \\
	\hline
$1$   & $1$ & $\beta$ & $0$ & $H$ & $G+J-i\epsp$  \\
$2$   & $-1+\beta$ & $0$ & $0$ & $G+H$ & $J-i\epsp$  \\
$3$   & $-\beta$ & $0$ & $0$ & $G$ & $J-i\epsp$  \\
$4$   & $-1$ & $\beta$ & $B$ & $C+E$ & $A+D+F-i\eps$  \\
$5$   & $1-\beta$ & $0$ & $A+B+C$ & $E+D$ & $F-i\eps$  \\
$6$   & $\beta$ & $0$ & $A$ & $D$ & $F-i\eps$  \\
	\hline
\end{tabular}
\label{coeff_T}
\ec
\end{table}

We have an important remark here. 
\Eq{T_result} is very different from Eq.~(6.26) of 
\cite{hooft_velt} in the respect of numerical evaluation. 
't Hooft and Veltman suggested a way to proceed with the latter 
by splitting the residue into two parts: one part has the same 
form as the first term in \eq{T_result} but with the variable 
$y$ in the argument of the \logas\ replaced by the pole of the 
corresponding previous factor, the other part is related to 
the \etaf . Eq.~(6.26) of 
\cite{hooft_velt} has two disadvantages. First, this equation 
is used in connection with the function $\sfun_3$ defined in Appendix~B 
of \cite{hooft_velt}. However, $\sfun_3$ is only well defined 
if $y_0$ is not a root of the argument of the second logarithm. 
In fact, there are several cases where the second logarithm 
of $\sfun_3$ is divergent but the final result of $\tf$ is finite. This 
point is made clear from the calculation of the function $\sfun_1$ 
in \appen{appen_T1}. Second, this equation leads to the evaluation of several \etaf s which 
might suffer from nasty cancellations. In particular, the \etaf s in 
eqs.~(6.26, B.2) of 
\cite{hooft_velt} may cause serious problems. 

The representation (\ref{detQ_abc}) leads naturally to two ways of numerical 
evaluation. This will be discussed in \sect{code}.

\subsubsection{Special cases: $A=0$ or $B=0$}
These special cases can occur when we apply \eq{d0_1zero} for the case 
of more than one lightlike momenta. 

If $B=0$ then both denominators in \eq{def_T} are linear 
in $y$ then the $y$-integral can be easily performed. The 
result reads
\bea
\tf(A,0,C,G,H;D,E,F,J)&=&\fr{1}{(HA-CG)(y_2-y_1)}\sum_{i=1}^2\sum_{j=1}^4(-1)^{i+j}\crn
&\times&\int_0^1dy\fr{1}{y-y_i}\ln(A_jy^2+B_jy+C_j),\label{T_B_sumij}
\eea
where $y_{1,2}$ are the two roots of 
\bea
(HA-CG)y^2+(HD-CJ-EG)y+H(F-i\eps)-E(J-i\epsp)=0
\eea
or 
\bea 
y_1&=&\fr{-(HD-CJ-EG)-\sqrt{\det(Q)}}{2(HA-CG)},\;  y_2=\fr{-(HD-CJ-EG)+\sqrt{\det(Q)}}{2(HA-CG)},\crn
\det(Q)&=&(HD-CJ-EG)^2-4(HA-CG)(HF-EJ).
\eea 
The discriminant is the determinant of the Landau matrix $Q$ if our calculation is set in the 
context of \eq{d0_1zero} or \eq{d0_T}. 
The other coefficients are given in \tab{coeff_T_B}.
\begin{table}[h]
\bc
\caption{The coefficients of the $\tf$ function}
\begin{tabular}{cccc}
	\hline
$j$ &  $A_j$ & $B_j$ & $C_j$ \\
	\hline
$1$   & $A+C$ & $E+D$ & $F-i\eps$  \\
$2$   & $0$ & $G+H$ & $J-i\epsp$  \\
$3$   & $0$ & $G$ & $J-i\epsp$  \\
$4$   & $A$ & $D$ & $F-i\eps$  \\
	\hline
\end{tabular}
\label{coeff_T_B}
\ec
\end{table} 

If $A=0$, we change the integration order by using \eq{int_x_y} and 
then change the integration variables in two successive steps: $x=x^\prime +y$ 
and $y=1-y^\prime$. At the end, one should obtain the 
following relation
\bea
\begin{aligned}
\tf(0,B,C,G,H; & D,E,F,J)=\tf(B+C,0,-C,-G-H,G;\\
& -D-E-2B-2C,D+C,D+E+B+C+F,G+H+J).
\end{aligned}
\label{T_A}
\eea
Thus, we get back to the above case of $B=0$.

We conclude this section by observing that the above results can be 
also applied for the case of real masses. In this case, the two poles 
$y_{1,2}$ in the rhs of eqs.~(\ref{d0_y_13_sumij}, \ref{T_result}, \ref{T_B_sumij}) can be real and the \iepsp\ is needed. The position of the 
poles depend on the value of internal masses and external momenta. Thus, 
it might happen that the two poles become equal and pinch the 
integration contour, $[0,1]$. If this occurs then the $D_0$ function can have 
a four-point leading Landau singularity (LLS). The issue of Landau singularities 
will be discussed in the appendices. 

\section{The code \doc}
\label{code}
The above method has been implemented into a Fortran 77 code, \doc , 
which can calculate the scalar one-loop four-point integral, $D_0$, with 
complex/real internal masses. The restriction is that at least one external 
momentum must be lightlike or $\alpha$ must be real (see \eq{cond_alpha}). The code is best exploited 
when included into the library \lt\ where a reduction procedure to calculate tensor
one-loop five-point integrals with complex masses is available. 
Thus, the new version of \lt\ (\verslt ) can handle 
all the one-loop integrals with complex masses up to five-point. 
The original code \doc\ can be downloaded at:
\bc
{\texttt http://wwwth.mppmu.mpg.de/members/ldninh/index.html}
\ec
\verslt\ including \doc \footnote{Thomas Hahn has adapted this code to \lt . 
The structure is unchanged. The way of coding is slightly modified. 
The numerical results of the two codes should be identical.} can be downloaded at: 
\bc
{\texttt http://www.feynarts.de/looptools/}
\ec

In the language of \lt , the $D_0$ function with complex masses, also named \doc , 
is invoked in the same way as the usual \dor\ with real masses. However, 
{\em all} the arguments of \doc\ must be of double complex type.  

The implementation of \eq{T_result} is straightforward. 
However, there is an important point 
we want to stress. The residue term on the rhs may cause 
serious problem of numerical cancellation when the imaginary part of 
the $R_i$s become very small. Thus we want to eliminate 
this term as far as possible. Our algorithm is based on the simple fact 
that this contribution vanishes if the pole is outside of the \betatr . 
If $\beta$ is real, the triangle becomes a line in the 
complex plane. The residue contribution vanishes. 
If $\beta$ is complex, there are two choices. First, we take 
$\beta=\beta_1$ which is the largest root of \eq{eq_beta}. 
There is a very efficient method to check whether a point (pole) is inside a triangle 
by using the barycentric technique. This idea is very simple. From 
a triangle, ABC, one can define a barycentric reference frame 
($\overrightarrow{AB}$, $\overrightarrow{AC}$) where A is the origin. The coordinates of 
a point P, ($x$, $y$), are easily calculated by writing:
\bea
\overrightarrow{AP}=x\overrightarrow{AB}+y\overrightarrow{AC}.
\eea
Multiply both sides by $\overrightarrow{AB}$ and $\overrightarrow{AC}$ to get 
a system of two equations to solve for ($x$, $y$). The conditions for P to be 
inside the triangle ABC are: $x$ and $y$ must be positive and $x+y\le 1$. 
However, one sees that the problem of numerical cancellation is still present 
if $x$ or $y$ is very small, \ie\  the point P is very close to the border of the triangle. 
If this happens, we repeat the above steps 
with the second choice $\beta=\beta_2$. That is what we have done in the code.\\
In the worst case, if $x$ or $y$ is very small for both choices 
of $\beta$ then the code will produce a warning of numerical cancellation but still continue the computation. The result in this case is not 
reliable and must be cross-checked. The code includes a means to do this. 
Indeed, we have written two different codes, called versions 1 and 2, 
to evaluate the rhs of \eq{T_result} based on two different strategies. The first one is described 
as above. In the second one, the residue contribution is always 
calculated as long as $\beta$ is complex. One has to specify which version when calling 
the main function:
\bc
{\texttt d0c=D0Cn(meth,p1,p2,p3,p4,p1p2,p2p3,m1,m2,m3,m4)}
\ec
where {\texttt meth=0 (version 1) or 1 (version 2)}. The two versions are different only 
when $\beta$ is complex. For \verslt , one can choose 
different versions by defining a version key before calling the \doc\ function as
\begin{center}
{\texttt call setversionkey(i*KeyD0C) \hspace*{3.3cm} }\\
{\texttt d0c=D0C(p1,p2,p3,p4,p1p2,p2p3,m1,m2,m3,m4)}
\end{center}
where \tti\ can be 0 (compute version 1), 1 (compute version 2), 
2 (compute both versions, return 1) or 3 (compute both versions, return 2). 
The default value is \tti =0.

The structure of the original code, written in double precision Fortran 77, is as follows.
\begin{itemize}
\item \dcn : the main function to return the value of the integral $D_0$ as 
a function of masses and momenta (see \eq{d0_def_ki}). It rotates 
the momenta to the right position and prepares input parameters for 
\eq{d0_xyz}.
\item \dcabc : a subroutine called in \dcn\ to calculate the rhs of \eq{d0_xyz}.
\item \ftot\ is a subroutine to calculate 
the function $D_0^{(13)}$  as defined by \eq{d0_2zero}.
\item \ftn (including \ftnlin\ to handle special cases) is a subroutine to 
calculate the $\tf$ function (defined by \eq{def_T0}) by using the 
barycentric technique described above. It belongs to version 1.
\item \ftnresd (including \ftnlin\ to handle special cases) is another 
subroutine to calculate the $\tf$ function for cross-checking purpose. 
It belongs to version 2.
\item \fsn\ and \frf\ are subroutines to calculate the functions $\sfun_{1}$ and 
$\rf_1$ defined in \appen{appen_T1}. 
\item \fstwo\ is a subroutine to calculate the functions $\tf$ and $D_0^{(13)}$ 
if the integrand is just a \loga , \ie\ the prefactor is just a constant.
\item Fundamental functions including Spence function, \loga , \etaf\ are taken 
from \ff .
\end{itemize}	
The code has been carefully checked as follows. Since the \iepsp\ is kept 
explicitly in our code, the code works also for the case of real masses. 
In this case, our results have been checked by comparing with the D0 function 
of \lt . Full agreement with the version 'b' of \lt\ has been found\footnote{\lt\ has two versions, 'a' and 'b', to calculate 
the D0 function \cite{looptools_5p}. Version 'a' is based on \ff\ while version 'b' uses the method of \cite{denner_d0}.}. 
We then introduced the widths for internal 
particles and checked the results in two limits: very small and very large widths. 
In the former case, we recovered the results of real masses. For the latter case, the results 
were checked against the ones of a routine where the integration 
is done numerically by using \bases\ \cite{bases}. We observed good agreement within the integration error (see \tab{num_com1}). 

Tables~\ref{num_com2},\ref{num_re1},\ref{num_re2} give some numerical results of the code for two cases: 
complex and real masses. For the latter we observed that our code agrees with 
version 'b' of \lt\ while disagrees with version 'a'. In fact, our code produces the same results 
with both versions of \lt\ in almost all physical cases. The results shown in Tables~\ref{num_re1},\ref{num_re2} 
are a few cases where disagreement has been found\footnote{\lt\ does produce some warning or error 
messages in these cases.}. 
Thus, care must be taken when using loop libraries 
and one should always cross-check the results by using different methods. For instance, it can be 
very useful to check the results of real masses in a tricky situation by using 
the code for complex masses with a very small imaginary part.

\begin{table}[th]
\small
\bc
\caption{Comparisons with a numerical integration routine using \bases , 
complex masses with large imaginary parts: $m_1^2=28900-25500i$, 
$m_2^2=6400-16000i$, $m_3^2=8100-22500i$, $m_4^2=40000-80000i$, $p_1^2=100$, $p_2^2=32400$, $p_3^2=400$
$p_4^2=250000$, $(p_2+p_3)^2=84100$. Numerical integration errors are about $(0.03\%, 0.07\%)$.}
\begin{tabular}{ccc}
	\hline
$(p_1+p_2)^2$ & \doc & Numerical integration \\
	\hline
48400 & (-1.50823957E-10,1.62902558E-11) & (-1.5077496E-10,1.63063608E-11) \\
52900 & (-1.51384299E-10,1.37108163E-11) & (-1.51339092E-10,1.37149562E-11) \\
57600 & (-1.51834347E-10,1.09709949E-11) & (-1.51787895E-10,1.09683749E-11) \\ 
62500 & (-1.52150627E-10,8.07814957E-12	) & (-1.52101762E-10,8.07660917E-12) \\ 
	\hline
\end{tabular}
\label{num_com1}
\ec
\end{table} 
%

\newpage
\begin{table}[th]
\bc
\caption{Some results of \doc , complex masses with small imaginary parts: $m_1^2=28900-255i$, 
$m_2^2=6400-160i$, $m_3^2=8100-225i$, $m_4^2=40000-800i$, $p_4^2=250000$, $(p_1+p_2)^2=72900$, 
$(p_2+p_3)^2=84100$}

\begin{tabular}{cccc}
	\hline
$p_1^2$ &  $p_2^2$ & $p_3^2$ & \doc$\times 10^{-10}$ \\
	\hline
100   & 32400 & 400 & (-37.6230835,-134.501524)  \\
0   & 32400 & 400 & (-37.8749852,-134.892905)  \\
0   & 32400 & 0 & (-38.6796394,-136.041222)   \\
0   & 0 & 0 & (-28.9137581,16.3570695)  \\
	\hline
\end{tabular}
\label{num_com2}
\ec
\end{table} 
%
\begin{table}[h]
\small
\bc
\caption{Comparisons between \lt\ and \doc , real masses: $m_1^2=m_2^2=m_3^2=m_4^2=30276$, 
$p_4^2=14400$, $(p_1+p_2)^2=57600$}
\begin{tabular}{ccc}
	\hline
$(p_1^2,p_2^2,p_3^2,(p_2+p_3)^2)$ & (0,124609,20,129630) & (0,0,20,73473) \\
	\hline
\lt\ 'a'$\times 10^{-10}$ & (-8.4106454,227.290811) & (19.8584112,-8.23191831) \\	
	\hline
\lt\ 'b'$\times 10^{-10}$ & (-8.4106454,26.7103536)  & (3.32391023,-6.755152E-16)  \\	
	\hline
\doc$\times 10^{-10}$ & (-8.4106454,26.7103536) & (3.32391023,-2.067952E-15)\footnotemark \\	
	\hline		
\end{tabular}
\label{num_re1}
\ec
\end{table}
\footnotetext{The 
imaginary part of $D_0$ is zero in this case since no normal-threshold-cut is open.}
\begin{table}[h]
\small
\bc
\caption{Continuation of \tab{num_re1}}
\begin{tabular}{ccc}
	\hline
$(p_1^2,p_2^2,p_3^2,(p_2+p_3)^2)$ & (0,0,0,734730) & (100,50,20,734730) \\
	\hline
\lt\ 'a'$\times 10^{-10}$ & (-0.400586866,-6.57420392) & (-0.401370928,-6.57507623) \\	
	\hline
\lt\ 'b'$\times 10^{-10}$ & (-0.400586866,1.85027075)  & (-0.401370928,1.85083346) \\	
	\hline
\doc$\times 10^{-10}$ & (-0.400586866,1.85027075) & (-0.401370928,1.85083346) \\	
	\hline		
\end{tabular}
\label{num_re2}
\ec
\end{table}
\section{Conclusions}
The Fortran code \doc\ allows to calculate the \df\ with complex masses for 
all physical processes at colliders. The method and structure of the 
code were explained. The results related to massless external particles and the discussion 
of Landau singularities complement the study of 't Hooft and Veltman. 
The code has been included in \verslt . 
Please refer to this 
paper if you use the code. Bug reports or comments should be sent 
to the authors\footnote{{\texttt Email: ldninh@mppmu.mpg.de} or {\texttt tdao@mppmu.mpg.de}}.

\vspace{0.5cm}
\noi {\bf Acknowledgments} \\
We thank F.~Boudjema for helpful discussions and for careful reading 
of the manuscript. A.~Denner is acknowledged for useful discussions and for 
checking some results with real masses. Special thanks go to T.~Hahn for 
valuable suggestions, comments and his careful reading 
of the manuscript. Indeed, he has spent a lot of time and 
effort to improve and adapt our code to \verslt , thereby making the code 
more useful for practical calculations. 
Figures were produced with JaxoDraw \cite{Binosi:2003yf}.     

\vspace{1cm}

\renewcommand{\thesection}{\Alph{section}}
\setcounter{section}{0}
\renewcommand{\theequation}{\thesection.\arabic{equation}}
\setcounter{equation}{0}
\noi {\Large {\bf Appendices}}

\section{Four-point leading Landau singularity}
\label{appen_real}
The method of this paper can also be applied for the case 
of real internal masses. In this case the $i\eps$-prescription 
is needed to define various multivalued functions (\logas , Spence functions) in the final results. 
From the practical viewpoint, this method is not 
as efficient as other methods based on projective transformation
\cite{hooft_velt, denner_d0} since the results are written in terms of 
a larger number of Spence functions. However, it has some useful virtues as 
a direct calculation and we will show that this method is indeed a very 
easy means to understand the analytical properties 
of the scalar four-point function. To illustrate this purpose, 
we first consider the following function
\bea
\tf_{1}=\int_0^1dy\fr{\ln(Ay^2+By+C-i\eps)}{Ky^2+Ly+N-i\epsp}.\label{def_T1}
\eea
An important property of this function 
is that the denominator has, in general, two poles 
\bea 
y_{1}&=&\fr{-L-\sqrt{L^2-4KN}}{2K}-\fr{i\epsp}{\Rel{\sqrt{L^2-4KN}}}, \crn
y_{2}&=&\fr{-L+\sqrt{L^2-4KN}}{2K}+\fr{i\epsp}{\Rel{\sqrt{L^2-4KN}}} .
\label{root_y12_eps}
\eea If these poles are real then 
it may occur that, 
in some region of parameter space, they pinch the integration 
contour $[0,1]$, leading to a singularity. This phenomenon occurs when 
the two roots become equal\footnote{This cannot happen in the case of complex masses since the 
discriminant, the Landau determinant $\det(Q)$, is protected from zero by an imaginary part.}, 
\ie\  $\det(Q)=L^2-4KN\to 0$. Since $y_2-y_1\propto 2i\epsp\neq 0$, we can split $\tf_1$ 
into two terms as usual
\bea
\tf_{1}=\fr{1}{K(y_2-y_1)}\int_0^1dy\left(\fr{1}{y-y_2}-\fr{1}{y-y_1}\right)\ln(Ay^2+By+C-i\eps).
\eea
Using the results of \appen{appen_T1} and taking the limit $y_2\to y_1=y_0=-L/(2K)$, $y_0$ is real, we get
\bea
\tf_{1}^{sing.}=\fr{\eta(i\eps_1,i\eps_1,\fr{i\eps_1}{y_0(y_0-1)})}{\sqrt{\det(Q)}}
\left[\ln(A-i\eps)+\eta(-z_1,-z_2)+\nf(z_1,y_0)+\nf(z_2,y_0)\right]
\label{limit_T1_full}
\eea
where $\eps_1=\epsp/\Rel{\sqrt{L^2-4KN}}$, $z_{1,2}$ are the roots of the argument of the logarithm and
\bea
\nf(z,y_0)&=&\left\{
\begin{array}{rl}
\ln(y_0-z) & \text{if } z\neq y_0\\
-\ln(-i\eps_z)-\pi i\sign(\eps_z) & \text{if } z = y_0 
\end{array} \right.\crn
\eta(a,b,c)&=&2\pi i[\theta(-\Img a)\theta(-\Img b)\theta(\Img c)-\theta(\Img a)\theta(\Img b)\theta(-\Img c)].
\eea
We remark that if 
\bea
y_0(y_0-1)<0, \hs \text{or}\hs 0<y_0<1
\label{cond_sign}
\eea
then the result is divergent. This is the origin of the four-point LLS. We also 
observe that if $y_0=z_{1}$ or/and $z_2$ then we have the coincident of four-point and three-point 
Landau singularities. The latter will be discussed in \appen{appen_T1}. 

The \df , $D_0$, is a sum of $\tf_1$s. The above conditions for leading Landau singularity are 
therefore necessary but not sufficient. In fact, the conditions for 
scalar one-loop integrals to have a LLS can be obtained by using 
Landau equations \cite{landau, Coleman:1965xm, ninh_thesis}. However, one might expect 
to find out the same conditions from a direct calculation. To illustrate the key points 
while keeping our explanation as simple as possible, we consider the case of 
$D_0$ function with two opposite lightlike external momenta, see \eq{d0_y_13_sumij}. 
We ignore the case of coincident singularity and assume $y_0\neq z_{1,2}$, 
\eq{limit_T1_full} becomes
\bea
\tf_{1}^{sing.}=\fr{\eta(i\eps_1,i\eps_1,\fr{i\eps_1}{y_0(y_0-1)})}{\sqrt{\det(Q)}}
\ln(Ay_0^2+By_0+C-i\eps),
\label{limit_T1_LLS}
\eea
where we recall that $y_0$ is real. Apply this result to \eq{d0_y_13_sumij} we get
\bea
D_0^{(13),sing.}=\fr{\eta(i\eps_1,i\eps_1,\fr{i\eps_1}{y_0(y_0-1)})}{\sqrt{\det(Q)}}
\left(\ln\fr{S_0+T_0-i\eps}{U_0+V_0-i\eps}-\ln\fr{S_0y_0+T_0-i\eps}{U_0y_0+V_0-i\eps}\right)
\eea
where the index $0$ in the rhs means that the functions are real (without $i\eps$) and 
calculated at $y=y_0$. We now use the following property
\bea
\fr{S_0y_0+T_0}{U_0y_0+V_0}=\fr{S_0+T_0}{U_0+V_0}=\fr{S_0}{U_0}=\fr{T_0}{V_0}=P_0
\eea
to get
\bea
D_0^{(13),sing.}=-\fr{\eta(i\eps_1,i\eps_1,\fr{i\eps_1}{y_0(y_0-1)})
\eta(i\eps\fr{P_0-1}{U_0+V_0},-i\eps\fr{P_0-1}{U_0y_0+V_0},i\eps\fr{P_0-1}{P_0(U_0+V_0)})
}{\sqrt{\det(Q)}}.
\eea
Thus, necessary and sufficient conditions for the function $D_0^{(13)}$ to have a LLS are
\bea
\left\{
\begin{array}{l}
\det(Q)=0\\
0<y_0<1\\
(U_0+V_0)(U_0y_0+V_0)<0\hs \text{and}\hs T_0V_0<0
\end{array} \right.
\eea
where $y_0$ is the root of the equation $SV-TU=0$. We have checked that 
this result agrees with the general conditions obtained by using Landau equations \cite{ninh_thesis}. 

\section{Basic integral and three-point Landau singularity}
\label{appen_T1}
We calculate the following integral
\bea
\sfun_{1}=\int_0^1dy\fr{\ln(Ay^2+By+C-i\eps)}{y-y_0+i\epsp},\label{def_s1}
\eea
where $A$ is real, while $B$, $C$ and $y_0$ may be complex, with the restriction 
that the imaginary part of the argument of the logarithm has always the same sign 
for $0\le y\le 1$, as indicated by the "$-i\eps$". The "$-i\epsp$" is only needed 
when $y_0$ becomes real. 

The numerator can be written as
\bea
\ln(Ay^2+By+C-i\eps)=\ln(A-i\eps)+\ln(y-z_1)+\ln(y-z_2)+\eta(-z_1,-z_2),
\eea
where $z_{1,2}$ are the two roots of the argument of the logarithm. 
We have
\bea
\sfun_{1}&=&\left[\ln(A-i\eps)+\eta(-z_1,-z_2)\right]
\ln\fr{y_0-1}{y_0}+\rf_1(y_0,z_1)+\rf_1(y_0,z_2),\label{result_s1}
\eea
where the $\rf_1$-function is defined as
\bea
\rf_1(y,z)=\int_0^1dx\fr{\ln(x-z+i\eps_z)}{x-y+i\eps_y}
\eea
with $\eps_{z,y}$ are infinitesimal. If $y=z$ and being complex, we use
\bea
\rf_1(y,z)=\fr{1}{2}\ln\left(\fr{1-z}{-z}\right)[\ln(1-z)+\ln(-z)].
\eea
If $y=z$ and being real, we then use
\bea
\ln(x-z+i\eps_z)=\ln(x-z+i\eps_y)-\eta(x-z+i\eps_z,\fr{1}{x-z+i\eps_y})
\eea
to get 
\bea
\rf_1(y,z)&=&\fr{1}{2}\ln\left(\fr{1-y}{-y}-i\eps_y\right)[\ln(1-y+i\eps_y)+\ln(-y+i\eps_y)+2\eta(i\eps_y,-i\eps_z,\fr{i\eps_y}{1-y})]\crn
&+&\eta(i\eps_y,-i\eps_z,\fr{i\eps_y}{y(y-1)})[\ln(i\eps_y)-\ln(-y+i\eps_y)].\label{R1_sing}
\eea
We observe that if $0\le y=z\le 1$ {\em and} $\eps_y\eps_z<0$ then integration contour is pinched and this singularity corresponds to 
the anomalous threshold of scalar three-point integral (also called the three-point Landau singularity) 
whose nature is logarithmic. Since the value 
of the \etaf\  is purely imaginary, the imaginary part of $\rf_1$ is divergent while 
there is a finite jump in the real part of $\rf_1$ at the singular point. 

If $z\neq y$ then 
the result reads
\bea
\rf_1(y,z)=\ln(y-z)\ln\left(\fr{1-y}{-y}-i\eps_y\right)+\rf(y,z)\label{R1_gen}
\eea
where the function $\rf(y,z)$ is given in Appendix~B of \cite{hooft_velt}. When 
calculating the rhs of \eq{R1_gen}, if the argument $(y-z)$ becomes negatively real then 
one has to introduce an infinitesimal imaginary part $i\eps_{yz}$. The 
result does not depend on the sign of this imaginary part as long as we introduce it 
everywhere in the rhs of \eq{R1_gen}. Thus, the simplest way is to write $y-z=y-z+i\eps_z$. 
This completes our calculation of $\sfun_1$.

\end{document}